\documentclass[%
 reprint,
 amsmath,amssymb,
 aps,
]{revtex4-2}

\usepackage{graphicx}% Include figure files
\usepackage{dcolumn}% Align table columns on decimal point
\usepackage{color}
\usepackage{marginnote}
\usepackage{slashed}
\usepackage{bm}
\newcommand{\tr}{\,\mathrm{tr}}
\newcommand{\Tr}{\,\mathrm{Tr}}

\usepackage{bm}% bold math
\usepackage{dsfont}
\usepackage[caption=false]{subfig}

\begin{document}

\preprint{APS/123-QED}

\title{Topological quantum criticality of the  disordered Chern insulator}

\author{Mateo Moreno-Gonzalez}
\affiliation{Institute for theoretical physics, Universität zu Köln, Zülpicher Str. 77a, 50937 Cologne, Germany}

\author{Johannes Dieplinger}
\affiliation{Institute for theoretical physics, Universit\"at Regensburg, 93040 Regensburg, Germany}

\author{Alexander Altland}
\affiliation{%
	Institute for theoretical physics, Universität zu Köln, Zülpicher Str. 77a, 50937 Cologne, Germany
}%

\begin{abstract}
We consider the two-dimensional topological Chern insulator in the presence of static disorder. Generic quantum states in this system are Anderson localized. However, topology requires the presence of a subset of critical states, with diverging localization length (the Chern insulator analog of the `center of the Landau band states' of the quantum Hall insulator.) We discuss geometric criteria for the identification of these states at weak disorder, and their extension into the regime of strong disorder by analytical methods. In this way, we chart a critical surface embedded in a phase space spanned by energy, topological control parameter, and disorder strength. Our analytical predictions are supplemented by a numerical analysis of the position of the critical states, and their multifractal properties.  

\end{abstract}

\maketitle

\section{Introduction}
Next to the one-dimensional SSH chain the quantum anomalous Hall insulator (aka Chern insulator) is the most basic topological insulator. Defined in two dimensions, the system does not possess symmetries other than hermiticity of the Hamilitonian (class A). It is `anomalous' in that its topological order is intrinsic, and does require an external magnetic field as in the  conventional quantum Hall insulator. (In fact, the first theoretical model of a AQH insulator was proposed by Haldane \cite{haldane1988model} as  a `field free' realization of a quantum Hall effect.)

In the presence of static disorder, the AQH band insulator turns into a topological Anderson insulator \cite{li2009topological,groth2009theory}, and the spectral band gap becomes a mobility gap. Crucially, however, at least some of its states  must remain delocalized. As with the critically delocalized states in the center of the IQH Landau levels, the presence of such states is required for consistency. If there were not any, one would end up with contradictions such as vanishing of the \emph{transverse} Hall conductance, whose integer quantization is the defining feature of this system \cite{prange1981quantized}. Alternatively one may reason that the band of conducting surface states cannot simply disappear but must hybridize with states from the other surface somewhere up in the spectrum via delocalized bulk states. 

\begin{figure}[ht]
    \centering
    \includegraphics[width=0.75\columnwidth]{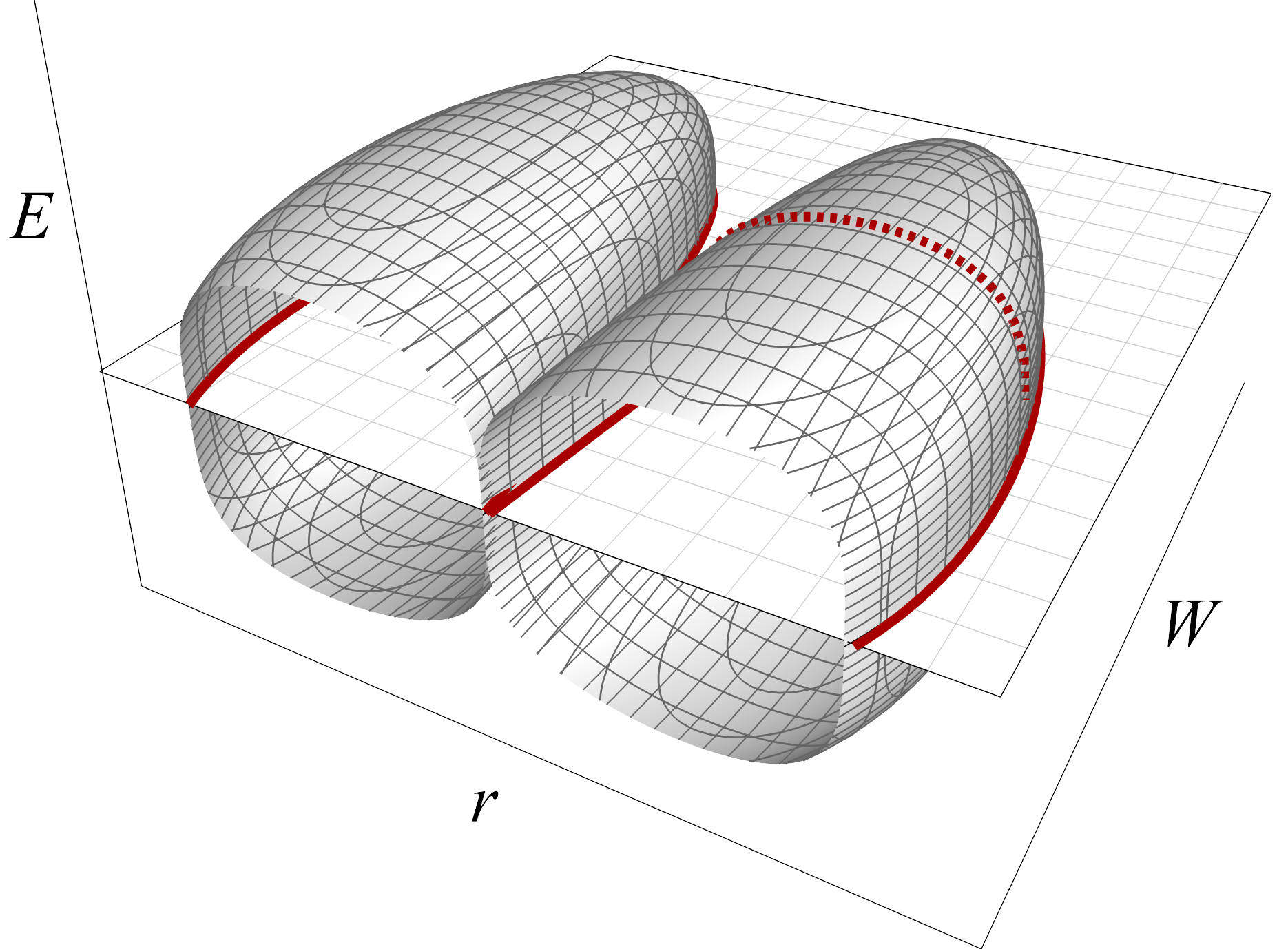}
    \caption{Schematic illustration of the critical surface of a Chern insulator. The clean system is defined in terms of different topological sectors realized as a function of a control parameter $r$, for example, $\mathrm{Ch}_\mathrm{1}=0\to 1\to -1\to 0$. For a given value of $r$ and disorder strength, $W$, delocalized states exist at a critical energies $E_l(r,W)$, one for each band. The merging of these bands at the center of the spectrum marks the stability boundary of the TI; for larger disorder the system has become a trivial Anderson insulator. (For systems with high Chern indices, $\mathrm{max}\,|\mathrm{Ch}_l|>1$, the destruction of topological order for increasing $W$ occurs in the successive merging of multiple surfaces $E_l(r,W)$.)}
    \label{fig:PhaseSurface}
\end{figure}

However, to the best of our knowledge the physics of AQH state delocalization has not
yet been systematically explored. Is there a single energy at which states delocalize
and if yes, where in the band is it situated? How much disorder is required to
destroy the delocalized states, and in this way the topological phase at large? What
are the critical properties of the delocalized states and how do they compare to
those of the IQH Landau level center states? In this paper, we address  these
questions within a two-thronged approach combining field theoretical constructions
with high performance direct diagonalization. Indeed it turns out that analytical
methods go a long way in the characterization of AQH delocalized states. The main result will be a phase diagram spanned by three parameters. The
first is a parameter, $r$, controlling the topological properties, i.e. the
topological index, $\nu(r)\in \mathbb{Z}$, of the clean system. The second is band
energy, $E$, and the third the disorder strength, $W$. Embedded in this parameter
space there lies a two dimensional surface, whose crossing implies criticality and
along with it the divergence of correlation lengths, or `delocalization', see Fig.\ref{fig:PhaseSurface} for a schematic. For
example, for $E=0$, the critical line $W(r)$, marks a stability boundary at which
disorder enforces a phase transition between the topological and a trivial insulator.
The critical value $E(r,W)$ defines the value in energy at which state delocalize for a
given control parameter and disorder strength, etc.

%%%%%%%%%%%%%%%%%%%%
%%%%%%%%%%%%%%%%%%%%

At weak  disorder,   the dependence $E(r,W)$ is encoded in the ground state  of the
clean topological insulator. (Specifically, we will show that for a simple two-band
model of the Chern insulator, the critical state is defined by a half-integer
Berry flux quantization condition.) For stronger disorder, the computation of the
critical energy gets a little more involved, but even then  requires no more than
knowledge of the SCBA Green function, i.e. $G(z)=(z-H(k)-\Sigma)^{-1}$, where $H(k)$
is the clean Hamiltonian, and $\Sigma$ the average quasiparticle self energy due to
impurity scattering. The condition for criticality then follows via computation of a
few integrals of Green functions over the Brillouin zone, which  may be done
analytically, or numerically.

The result of this analysis is a comprehensive phase portrait of the disordered Chern insulator, including predictions for its stability boundaries, the critical states, and conduction properties at finite system sizes before Anderson localization becomes effective in the thermodynamic limit. 

In the next section, we summarize the main results of our study. This will be followed by section \ref{sec:derivation_action} where we present a self contained derivation of the topological field theory underlying our analytical results. In section \ref{sec:results} we discuss the results obtained from both, the numerical and analytical calculations and we conclude in section \ref{sec:conclusions}.

\section{Summary of Results}
\label{sec:Summary}

We begin this section with a short review of the clean Chern insulator. This will be
followed by the discussion of an effective field theory describing the critical
physics of the disordered system, and a comparison to the conventional quantum
Hall insulator. We conclude with a discussion of numerical results.

\subsection{Clean Chern insulator} % (fold)
 \label{sub:clean_chern_insulator}
 
The clean two-dimensional (more generally, even-dimensional) Chern insulator is a
topological insulator characterized by non-vanishing Chern number, $\mathrm{Ch}_l$,
carried by individual of its bands, $l=1,\ldots, N$. These numbers are constructed
from the Berry connection $a=i\langle k|dk \rangle=i \langle  k |\partial_{k_i}
k\rangle dk_i$, where $\{|k\rangle\}$ are the Bloch eigenstates of a specific band,
and $k=(k_1,k_2)$ is  crystal momentum \cite{berry1984quantal,thouless1982quantized,simon1983holonomy}. Integration of the Berry curvature $b=da= i
\langle \partial_{k_j} k |\partial_{k_i} k\rangle dk_i\wedge dk_j$ over the Brillouin zone then yields the
Chern number as,
$\mathrm{Ch}_l=\frac{1}{4\pi}\int_{\mathrm{BZ}} b$.

The simplest model system realizing a nontrivial set of Chern numbers is Haldane's two-band QAH insulator \cite{haldane1988model,kane2005quantum}, defined by the Hamiltonian
\begin{align}
    \label{eq:Hamiltonian}
    H&=\sin k_1 \sigma_1+ \sin k_2 \sigma_2+(r-\cos k_1-\cos k_2)\sigma_3\cr 
    &\equiv  \sum_{a=1}^3 h_a(k) \sigma_a,
\end{align}
where we have set the hopping strength to unity for simplicity. 
For $r\not=-2,0,2$ this is a two-band insulator with `valence' and `conduction' band Chern numbers $(\mathrm{Ch}_\mathrm{v},\mathrm{Ch}_\mathrm{c})=(0,0),(-1,1),(1,-1),(0,0)$ for $r<-2,-2<r<0,0<r<2,2<r$, respectively. In this particular case, the integral over the Berry curvature assumes the form $\mathrm{Ch}_\mathrm{c}=-\mathrm{Ch}_\mathrm{v}=  \frac{1}{4\pi}\int d^2 k \,n\cdot (\partial_{k_1}n\times \partial_{k_2}n))$, where $n=h/|h|$, i.e. it probes the number of windings of the Pauli vector $h=\{h_a\}$ defining the Hamiltonian over the unit sphere.

\subsection{Disordered Chern insulator} % (fold)
\label{sub:disordered_chern_insulator}

In the presence of translational invariance breaking disorder, the Bloch eigenstates
of the system get replaced by states which generically are Anderson localized. For
impurity scattering rates, $\tau^{-1}$, comparable to the band gap of the clean
system, the spectral gap gets filled by impurity states, with a
globally gapless spectral density. The topological properties of the system, too, are no longer characterized by momentum space invariants but instead by a effective field theory defined in real space.

To define this theory, consider the matrix $Q=T\tau_3 T^{-1}$, where $T \in
\mathrm{U}(2R)$ is a unitary matrix, $R$ a number of replicas ($R\to 0$, eventually),
and the doubling factor $2$ required to distinguish between retarded and advanced
Green functions in the computation of observables in this formalism. The matrix
$\tau_3 = \tau_3 \otimes \mathds{1}_R$ acts in this space, and its presence implies
that $Q\in \mathrm{U}(2R)/\mathrm{U}(R)\times \mathrm{U}(R)$ is element of a coset
space in which matrices commutative with $\tau_3$ are divided out. 

Promoting $Q=Q(x)$ to a matrix field defined in two-dimensional space, we note that there are just two rotationally invariant gradient operators. They define an effective action as
\begin{align}
    \label{eq:QAction}
    S[Q]= \int d^2 x \left(g\,\mathrm{tr}( \partial_i Q \partial_i Q) + \frac{\theta}{16\pi}\epsilon_{ij}\, \tr(Q\partial_i Q \partial_j Q)\right),
\end{align}
where the two coupling constants $g=\sigma_{xx}/8$ and $\theta=2\pi\sigma_{xy}$ are determined by the system's longitudinal and transverse Hall conductance, respectively (in units of the conductance quantum).

The second term in Eq.~\eqref{eq:QAction} is topological in nature. For a fictitious infinitely extended (or boundary-less) system it computes ($16\pi i \times$) the number of times $Q(x)$ covers the coset space $\mathrm{U}(2R)/\mathrm{U}(R)\times \mathrm{U}(R)$  as a function of $x$. In its absence fluctuations of the $Q$-matrix field will lead to a logarithmically slow downward renormalization of   $g\sim \sigma_{xx}$ --- two-dimensional Anderson localization \cite{abrahams1979scaling}. The same happens for generic values of  $\theta\sim \sigma_{xy}$. In fact, Pruisken \cite{pruisken1984localization,levine1984theory} derived the above action as the effective theory underlying  Khmelnitskii's two parameter scaling paradigm \cite{khmelnitskii1983quantization} according to which the flow of coupling constants $(g,\theta)\to (0,2\pi n)$ generically ends in localizing fixed points, $g=0$, with integer quantized Hall conductance, $\sigma_{xy}=n$.

The exceptional situation occurs at $\theta=\pi$. For this critical value the flow ends in a fixed point with finite conductance, $(g^\ast,\pi)$, the quantum Hall critical point. (The  critical point itself is described by an effective theory different from Eq.~\eqref{eq:QAction}. While the  identity of that theory remains unknown to date, our focus here is on the identification of criticality, and for that purpose Eq.~\eqref{eq:QAction} remains the appropriate diagnostic.)

The identification of the phases supported by a
two-dimensional class A topological insulator thus amounts to the computation of the
parameters $(g,\theta)$ for a given microscopic system description. Previous work \cite{ostrovsky2007quantumcriticality}
performed this task in a Dirac or `$k\cdot p$' approximation, valid for energies
close to a band closing point. Since, however, the critical surfaces may be buried
deeply in the Chern bands, this approximation is not an option here, we need to work
with the full lattice dispersion. Interestingly, abandoning the Dirac linearization
turns out to be a blessing, including from a computational perspective: the
construction of effective field theories building on top of a microscopic Dirac
Hamiltonian is met with ultraviolet singularities. These need to be regularized by
one of various available schemes, which, however, are all \emph{ad-hoc} from a condensed matter
perspective \cite{pauli1949regularization,bollini1972dimensional,veltman1972regularization}. However, these singularities are a mere artifact of the linearization,
they do not occur within for the full system with its bounded dispersion
relations. Relatedly, they obscure the geometric interpretation of the topological
angle as an integral over the Brillouin zone.

In section \ref{sec:derivation_action}, we will see that the derivation of the effective theory building on the lattice Hamiltonian is both more general and conceptually simpler than the Dirac approach. Specifically, it yields the topological angle at weak disorder of the two-band insulator as an integral
\begin{align}
    \label{eq:TopAngleWeakDisorder}
    \theta=\frac{1}{2}\int_\mathrm{\epsilon_k>E} d^2k  \,n\cdot (\partial_{k_1}n\times \partial_{k_2}n)).
\end{align}
While this expression is derived for the simplest model of a Chern insulator, the generalization to others is obvious: the topological angle is defined by the fraction of the full Berry flux $2\pi l$ carried by all states in the band above the reference energy $E$ (Since only $\theta \,\mathrm{mod}\,2\pi$ matters, we may equally compute the flux of states below $E$). Criticality occurs for states for which $\theta(E)=\pi$. 

Far from the weak disorder the fractional Berry flux becomes statistically distributed, and its mean value features as the topological angle $\theta$. This value can be represented as the sum of two momentum space integrals over SCBA broadened Green functions, aka the Smr\v{c}ka-St\v{r}eda coefficients \cite{smrcka1977transport}, $\theta^{\text{I}} = 2\pi \sigma_{xy}^{\text{I}}$ and $\theta^{\text{II}} = 2\pi \sigma_{xy}^{\text{II}}$.  Odd integer values $\theta=\theta^\mathrm{I}+\theta^{\mathrm{II}}=(2n+1)\pi$ serve as markers for topological quantum criticality, as in the IQH context. 

\subsection{Numerical analysis of the multifractal spectrum}
\label{sec:numerics}
\label{sec:multifrac}

Complementing the analytical approach, we diagnose criticality numerically via multifractal analysis of states at the critical points \cite{evers2008anderson,mirlin2000multifractality,rodriguez2011multifractal,puschmann2021quartic}. Our starting point is the tight-binding model in Eq. \eqref{eq:Hamiltonian} with periodic boundary conditions and on-site uncorrelated Gaussian disorder with width $W$. 
For a given wave function $\psi_E$ at energy $E$, the key objects of interest are the $q$-th moments
\begin{equation}
  \label{eq:InvParticipationRatio}
	P_q^E=\sum_{ij}^{N} \sum_{\alpha} \mid \psi^E_{ij,\alpha}\mid^{2q},
\end{equation}
where  $\psi^E_{ij,\alpha}$ labels the wave function element at lattice site $(i,j)$ and internal degree of freedom $\alpha=1,2$.

Th $q$-th of these moments scales with linear system size $N$ as 
\begin{equation}
	P_q^E(N)\propto N^{\tau_q},
	\label{eq_moments}
\end{equation}
where $\tau_q$ is the effective dimension. Extended metallic wave
functions in a $d$-dimensional lattice have dimension $\tau^\text{metal}_q=d(q-1)$, while  localized wave function show system size independent scaling $\tau^\text{loc}_q=0$. Finally, the fluctuations of a critical wave function are captured by  the anomalous part of the effective dimension \cite{evers2008anderson,rodriguez2011multifractal},
\begin{equation}
	\Delta_q^E=\tau_q^E +d(q-1).
\end{equation}
Specifically, the multifractal dimension of the quantum Hall transition has an
approximately parabolic spectrum with $\Delta_q^\text{QH}\approx0.25\,q\,(q-1)$\cite{evers2008multifractality}. Below, we will use this scaling as a benchmark for diagnosing quantum Hall
criticality in the Chern insulator. 

In practice we calculate the quantity 
\begin{equation}
	\tilde{\tau}_q(N;E,W,r)=\frac{\log \langle P_q^{E,W,r}(N) \rangle_\text{dis,E}}{\log N},
\end{equation}
where $\langle \cdot \rangle_\text{dis,E}$ denotes the double average over a small
energy window, consisting of $N_E$ subsequent wave functions as well as over
$N_\text{avg}$ disorder configurations. In the limit of infinite system size,  $\lim_{N \to \infty}
\tilde{\tau}_q(N) =\tau_q$. 

\begin{figure}[t!]
	\includegraphics[width=\linewidth]{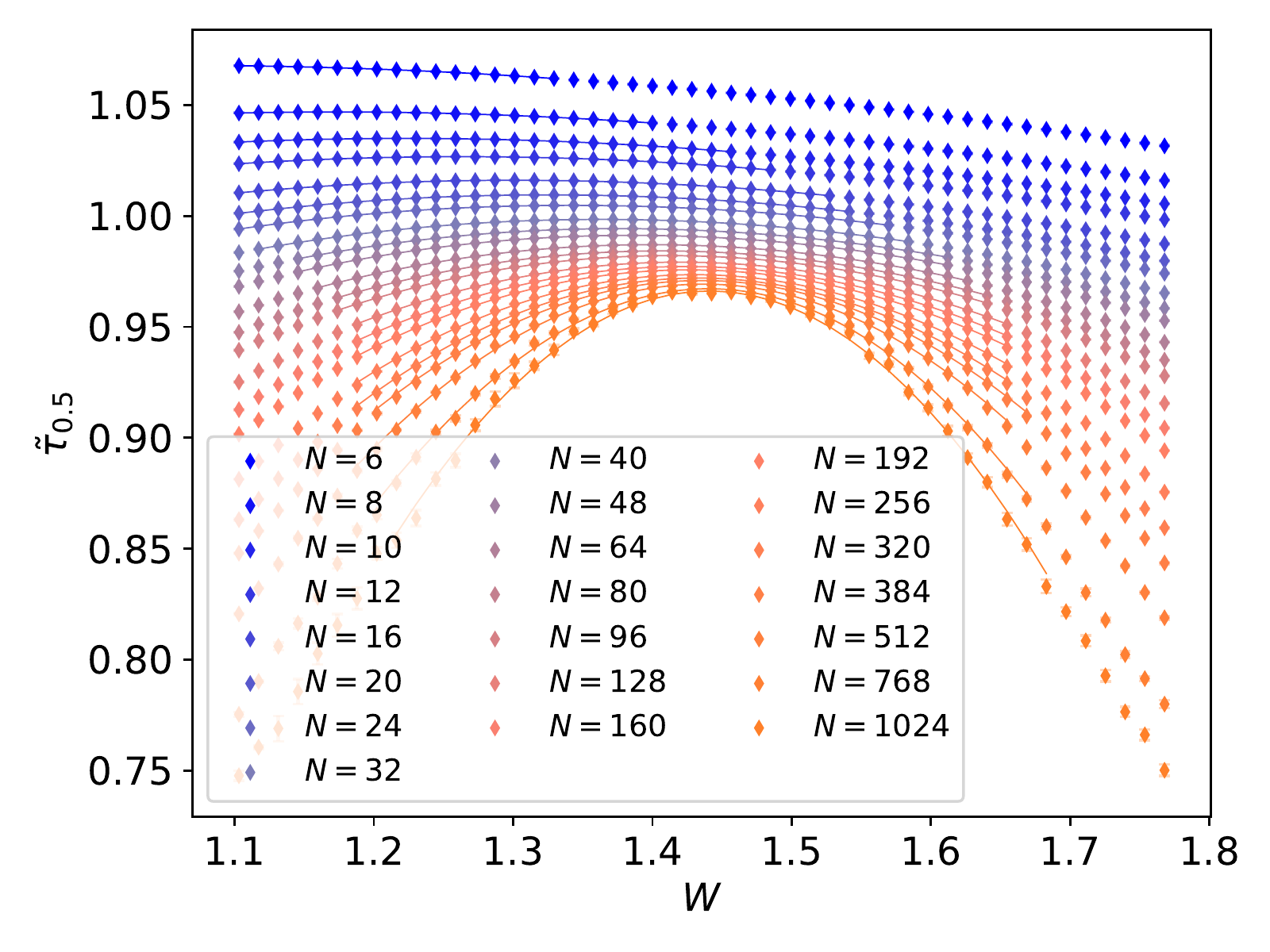}
	\caption{Effective dimension $\tilde \tau_q(E,W,r)$ for different system sizes at the point $(r,E)=(1.2,0.6)$ as a function of $W$. The data is approximately described by parabolas with increasing opening angle for increasing system size. The fit (solid lines) is done using a cubic polynomial Eq. \eqref{eq_app_fit_para} to account for asymmetries  away from the critical point.  }
	\label{f_app2}
\end{figure}

In Fig. \ref{f_app2} we show $\tilde{\tau}_{0.5}$ close to the critical surface for
fixed $r,E$ for different $W$. The critical state is identified as the maximum of the
data for a given system size $N$.  Finite size scaling and extrapolation to
$N\to\infty$ allows us to extract the limiting effective dimension $\tau_q$,  the
critical disorder strength at this system parameters, and the localization length
exponent $\nu$. Referring for details of the extrapolation procedure to
Appendix \ref{app_num}, we also extract the  corresponding irrelevant scaling
corrections $y_\tau, y_\nu$ from the curvature of the data for  different system
sizes.

\begin{figure}[t!]
	\includegraphics[width=\linewidth]{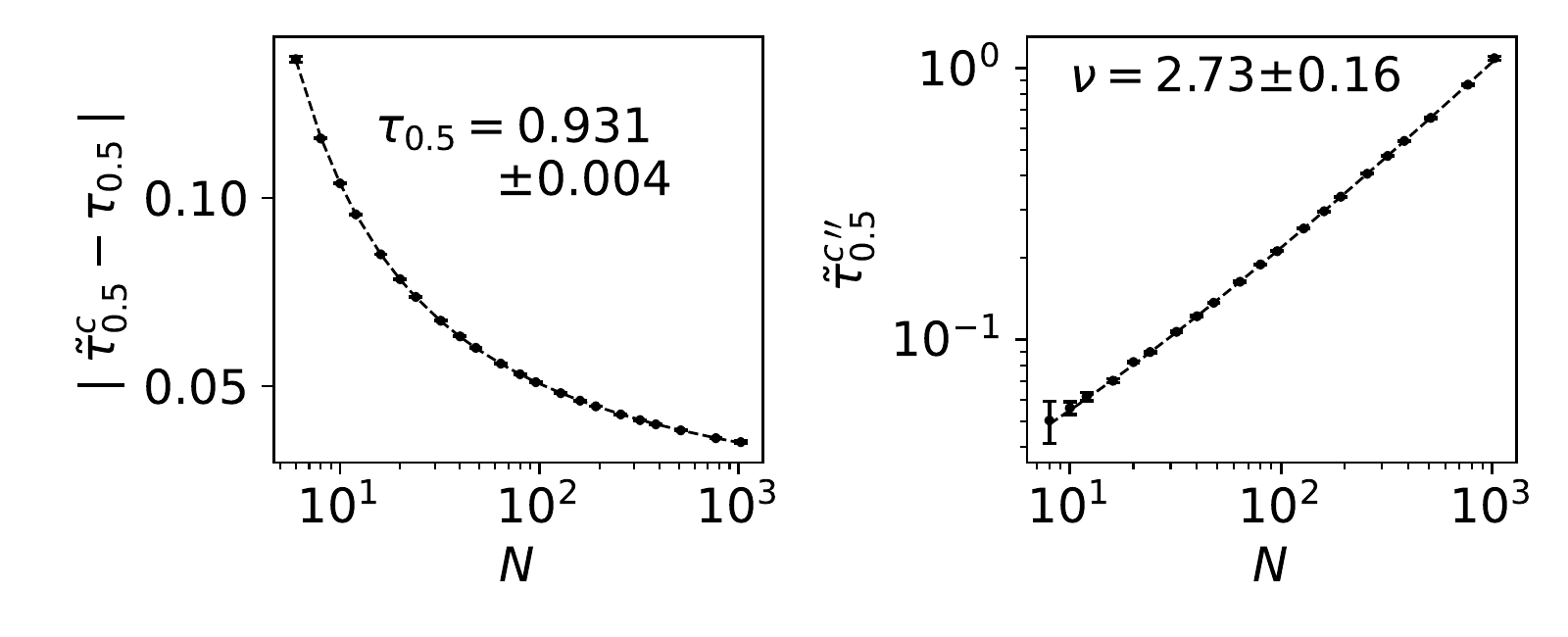}
	\caption{{Critical exponents of the topological phase transition. left:
			Using the fit function Eq. \eqref{fit_tau}, the effective dimension $\tilde{\tau}^c_q$, 
			extracted from the maxima of the curves in Fig. \ref{f_app2} converges as a function of the system size $N$ to $\tau_q=0.931\pm 0.004$.} 
		(right) system size scaling of the curvature $
		\tilde{\tau}^{c\,\prime\prime}_q$ of the fitted curves at the maxima in Fig.
		\ref{f_app2} fitted by Eq. \eqref{fit_nu}. The localisation length exponent is
		estimated to be $\nu=2.73\pm0.16$. The irrelevant exponents are shown in App. \ref{app_num}.}

	\label{f_app3}
\end{figure}

Fig. \ref{f_app3}  shows the scaling of the extrapolated critical exponents.
We find $\tau_{0.5}=0.931\pm0.004$,  $\nu=2.73\pm0.16$. Within the error bars this is consistent with results for  the integer quantum Hall effect in  Chalker-Coddington networks and tight binding models \cite{evers2001multifractality,puschmann2019integer,puschmann2021green,dresselhaus2022scaling}. 

The above discussion shows that field theory and numerical analysis provide complementary means to identify and analyze the critical state of the Chern insulator. In particular, one can be used to test the accuracy of the other. (We will see, that this validation goes in both directions.) Referring to section \ref{sec:results} for the detailed comparison, we will identify the position of the extended state in the conduction band given $r$ and $W$, and on this identify a phase boundary in the $r-W$ at the band center $E=0$. Building on this information, we will then map out the  full phase diagram in $(r,W,E)$ space. However, before turning to this comparison, we include a self contained discussion of the field theoretical apparatus underlying our results. This section may be skipped by readers primarily interested in results. 

\section{Derivation of the effective action} % (fold)
\label{sec:derivation_action}

In this  section we discuss the  derivation of the effective action
Eq.~\eqref{eq:QAction} of the disordered Chern insulator. While the initial steps
of the construction are standard, they are included here to keep the discussion self
contained. Emphasis will then be put on the derivation of the topological action, which is technically novel.

\subsubsection{Disorder average and stationary phase analysis} % (fold)
  \label{ssub:disorder_average_and_stationary_phase_analysis}
  
Our starting point is the Gaussian integral
\begin{align}
    Z=\int D\psi \, e^{-i\bar \psi(E+i\delta \tau_3-H-V)\psi},
\end{align}
where $\psi=\{\psi^{a}(x)\}$, $a=(s,r)$ is a $2R$ component Grassmann field, where $s=\pm$ distinguishes between retarded and advanced Green functions (corresponding to the Pauli structure $i\delta \tau_3)$, and $r=1,\ldots R$ is a replica index. The  Hamiltonian, $H$, represents the two-band model Eq.~\eqref{eq:Hamiltonian}, and $V=V(x)\sigma_0$ is potential disorder, where $\sigma_0$ is the unit matrix, and $\langle V(x)V(x')\rangle\equiv \tfrac{W^2}{2}\delta(x-x')$ Gaussian correlated with zero mean. 

From this functional, observables such as transport coefficients can be represented
as Gaussian integrals via the introduction of suitable source terms (with an implied
replica limit $R\to 0$). However, for our purposes, it will be sufficient to consider
the functional as it is. Following standard protocol, we average the functional over
disorder, to obtain a quartic term $\frac{W^2}{4}(\bar \psi \psi)^2$ in the action.
Decoupling the latter by a Hubbard-Stratonovich transformation and integration over
the fermion field leads a functional $Z=\int DA\,\exp(-S[A])$ with action
$\frac{1}{W^2}\int dx \tr A^2 - \tr \ln(E+i\delta \tau_3 - H - A)$, where
$A=\{A(x)^{ab}\}$ is a matrix field
\cite{altland1999supersymmetry,efetov1999supersymmetry,altland2010condensed}. A variation of the action in this field, leads to the stationary phase
equation $A(x)= \tfrac{W^2}{2}\tr (E+i\delta \tau_3 - H - A)^{-1}(x,x)$, i.e. a self
consistent Born equation with `impurity self energy', $A$. We parameterize a
matrix-diagonal solution compatible with the symmetry of the causal increment
$i\delta \tau_3$ as $A\to \Delta E + i \kappa \tau_3$. Here, $\Delta E$ and $\kappa$ are the quasiparticle energy shift and pole broadening induced by scattering, respectively.   Referring for a detailed discussion of
these parameters to Appendix \ref{sec:SCBA}. We note that for
infinitesimal $\delta$, the equation admits a continuous manifold of solutions,
$A=\Delta E + i \kappa Q$, where $Q=T \tau_3 T^{-1}$ with unitary $T\in
\mathrm{U}(2R)$ parameterizing the coset space mentioned in the previous section.

Substituting these configurations into the action, and upgrading the constant $T$ to a slowly varying Goldstone mode $T(x)$, we are led to consider the effective action 
\begin{align}
      S[T]=-\mathrm{tr}\ln\left(\sum\limits_{\mu} v_{\mu}(k)\sigma_\mu+i\kappa T(x)\tau_3 T^{-1}(x)\right),
  \end{align}  
where we absorbed the energy shift $\Delta E$ into $E$, neglected the infinitesimal $\delta$ in comparison to $\kappa$, and introduced the four component vector $v_\mu$ with $v_0=E$, and $v_a=-h_a$, $a=1,2,3$. In the following we expand this action in slow $T$-fluctuations, first leaving the detailed form of the momentum-dependent coefficients $h_a=h_a(k)$ unspecified. 

\subsection{Gradient expansion} % (fold)
\label{sub:gradient_expansion}

We begin our analysis of the fluctuation action with a unitary rotation of the tr ln, leading to 
\begin{align}
      S[T]=-\mathrm{tr}\ln\left( v_{\mu}\sigma_\mu+i\kappa \tau_3 +[T^{-1},v_\mu \sigma_\mu] T)\right),
  \end{align}  
  where a summation convention is applied, and the argument-dependence
  $v_\mu=v_\mu(k)$ and $T=T(x)$ is left implicit. Previous work performed this
  analysis for an effective Dirac Hamiltonian, $v_\mu=(E,-k_1,-k_2,m)$, for which the
  transformation of the logarithm is not innocent: It generates  the
  chiral anomaly, and the need for UV regularization. Here, we need not worry, as we
  are working with a manifestly UV regular theory.

 Assuming variation of the fields $T$ over scales much larger than the lattice
 spacing, we approximate the commutator up to second order in derivatives as
 \begin{align*}
  [T^{-1},v_\mu\sigma_\mu]T\simeq F_i \Phi_i - \tfrac{1}{2}J_{ij}\Phi_i \Phi_j,  
 \end{align*}
 with $F_i=i \partial_i v_\mu\sigma_\mu$, $J_{ij} =  \partial_i \partial_j  v_\mu\sigma_\mu $,  $\Phi_i =
 (\partial_i T^{-1}) T$, and the abbreviated notation $\partial_i v = \partial_{k_i}v$ and $\partial_i T = \partial_{x_i}T$.  Our task now is to evaluate the formal second order
 expansion 
  \begin{align}
    \label{eq:actexpansion}
      S[Q]&=-\tr \ln \left(v_\mu\sigma_\mu+ i \kappa \tau_3 +  F_i \Phi_i -
      \frac{1}{2}J_{ij}\Phi_i \Phi_j \right)\\
      &=-\underbrace{\tr  (GF_i \Phi_i - \frac{1}{2}G J_{ij}\Phi_i \Phi_j)}_{S^{(1)}}+\underbrace{\frac{1}{2}\Tr(GF_i \Phi_i)^2}_{S^{(2)}},\nonumber
 \end{align}

 with the  Green function 
 \begin{align}
    \label{eq:SCBAGreenFunction}
      G&=(i\kappa \tau_3 + v_\mu\sigma_\mu)^{-1}=D(i\kappa \tau_3+v_\mu\sigma^\mu),\cr &D=[(i\kappa\tau_3+E)^2 - h_a h_a]^{-1},
  \end{align} and the convention $x_\mu y^\mu=x_0 y_0 - x_a y_a$. In the following,
   we discuss how the two terms above yield the sum of a gradient term, and a
   topological term for the effective action of the system. Both contributions are of
   second order in derivatives, the difference being is that one contains
   $\partial_i
\partial_i$ derivative combinations, the other $\epsilon_{ij}\partial_i \partial_j$. 

\subsection{Topological action} % (fold)
\label{sub:topological_action}

In the construction of the topological action, we go fishing for antisymmetric
derivative combinations $\epsilon_{ij}\partial_i \partial_j$. As we show in Appendix \ref{sec:derivation_of_the_topological_action} these emerge from both  terms $S^{(1)}$ and $S^{(2)}$. On this basis, we obtain the topological action 
\begin{align}
    \label{eq:TopologicalAction}
    S_\mathrm{top}&=S_\mathrm{top}^{(1)}+S_\mathrm{top}^{(2)}=  \frac{\theta_2+\theta_1}{16\pi}\int dx \,\mathcal{L}_\mathrm{top}(Q),\cr 
    &\mathcal{L}_\mathrm{top}(Q)\equiv \epsilon_{ij}\tr(Q \partial_i Q \partial_j Q),
\end{align}
with coupling constants
\begin{align}
\label{eq:TopologicalCouplingConstants}
\theta_1&=  8\kappa \int (dk) D^+ D^-F_k,\cr 
\theta_2&=4\pi i\int (dk)\int\limits_E^\infty d\omega \, (D_{\omega}^{+2}-D_{\omega}^{-2}) F_k
\end{align}
where $(dk)=\frac{dk_1 dk_2}{(2\pi)^2}$ and 
\begin{align}
       \label{eq:Fk_def}
        F_k=\epsilon_{abc}h_a \partial_1 h_b \partial_2 h_c=(\partial_1 h\times \partial_2 h)\cdot h.
\end{align}
(Following standard conventions \cite{smrcka1977transport,pruisken1984localization}), we associate the one/two derivative action $S^{(1/2)}$ with the two/one contribution to the topological action, $\theta_{2/1}$.) 
 Our final task thus is to compute the coefficients $\theta_{1,2}$.
These integrals are straightforward for weak disorder, under a presumed hierarchy of energy scales 
\begin{align}
    \label{eq:EnergyHierarchy}
    \kappa \ll E\lesssim 1.
\end{align}
We first represent the propagators $D^s$ as
\begin{align*}
    D^s=\frac{1}{E^2-\epsilon^2+is \gamma \tau_3},\qquad \gamma=2\kappa E,\quad \epsilon^2=\sum_a h_a^2.
\end{align*}
Under the stated conditions, this leads to the approximation
\begin{align}
    \label{eq:DeltaApprox}
    \kappa D^+D^-=\frac{\kappa}{(E^2-\epsilon^2)^2+\gamma^2}\simeq \frac{\pi}{4E^2}\delta(E-\epsilon),
\end{align}
where here and throughout, $\epsilon>0$ is the positive root of $\epsilon^2$. Thus, 
\begin{align*}
    \theta_1\simeq \frac{2\pi^2}{E^2}\int (dk)F_k\delta(E-\epsilon),
\end{align*}
which is an on-shell integral probing the density of states at $E$. 
Turning to $\theta_2$, we note $D^{s2}_\omega\simeq \partial_{\omega^2}D^s_\omega$, and
\begin{align*}
    D^+_\omega-D^-_\omega\simeq -\frac{\pi i}{\omega} \delta(\omega-\epsilon)
\end{align*}
Entering with these relations into the integral defining $\theta_1$ and integrating by parts, it is straightforward to verify that
\begin{align*}
    \theta_2\simeq 2\pi^2\int (dk) F_k\left(\frac{\Theta(\epsilon-E)}{\epsilon^3}-\frac{\delta(\epsilon-E)}{\epsilon^2}\right).
\end{align*}
We note that the second, on-shell term cancels against $\theta_1$. To understand the
meaning of the first, recall that $\epsilon=|h|$. We may thus define the unit sphere
area element $S\equiv F/\epsilon^3=n\cdot (\partial_1 n \times \partial_2 n)$ with
unit vector $n=h/\epsilon$. Tidying up, we obtain the topological angle as in
Eq.~\eqref{eq:TopAngleWeakDisorder}.

\subsection{Gradient action} % (fold) \label{sub:gradient_action}

The gradient term of the action is obtained by similar inspection of $S^{(1,2)}$, this time focusing on derivative combinations of the form  $ \sim
F_i F_i$. As detailed in Appendix \ref{sec:derivation_of_the_gradient_action}, this leads to
\begin{align}
\label{eq:GradientAction}
    S_{\mathrm{grad}} &= I \int dx \,\tr \left( \partial_i Q \partial_i Q \right), \cr &I = I_{+}+ I_{-}+I_{+-},
\end{align}
with coupling constants given by
\begin{align}
    \label{eq:GradientCouplingConstants}
        &I_{+-} = \frac{1}{2}\sum_a\int (dk) (E^2+\kappa^2-\epsilon^2 + 2h_a^2) D^+ D^- \partial_i h_a \partial_i h_a,\cr 
    &I_\pm  = \frac{1}{4} \sum_a\int (dk) ((E+i \pm  \kappa)^2 - \epsilon^2 + 2 h_{a}^2 ) D^{\pm 2} \, \partial_i h_a \partial_i h_a.
\end{align}
We are left with the task to do the momentum integrals. For weak disorder, these integrals are analytically doable, if somewhat tedious. As a result, detailed in Appendix \ref{sec:derivation_of_eq_}, we obtain
\begin{equation}
     I = \frac{E^2-m^2}{2|E|\kappa} \Theta(E^2-m^2),
     \label{eq:intI}
\end{equation}
where $m=(r-c)$ and $c=2,0,-2$ depending on the Dirac cone around which we approximate. 

This result states that for weak disorder, diffusive quasiparticle propagation is limited to energies above the clean insulator band gap, $m$. For energies $E\gg m$, the coupling constant asymptotes to $\sim E/\kappa$ which is the characteristic scale for the conductivity of a weakly disordered two-dimensional conductor.

\subsection{Beyond the weak disorder limit}
\label{sec:beyondweakdis}

From  Eq.\eqref{eq:TopologicalCouplingConstants} we may calculate angle $\theta$ for arbitrary two-band  Hamiltonians $H$ and for values ($r$, $E$, $W$) such that the self consistent Born approximation underlying our theory remains valid, $E\kappa \gtrsim 1$. However, outside the weak disorder regime $E\kappa\gg1$ considered in the previous section the analytical computation of the integrals becomes cumbersome, or even impossible. 

Progress can nevertheless be made, starting from the following representation of the coupling constants in terms of energy/momentum integrals:
%\begin{widetext}
    \begin{align}
          \label{eq:sigmaxy}
           \theta_1 &= -\frac{i\pi}{2} \int (dk) \, \epsilon_{ij} \tr\left(\tau_2 G_E \partial_i G_E^{-1}\tau_1 G_E \partial_j G_E^{-1}\right),\cr
             \theta_2 &= \frac{\pi \epsilon_{\alpha \beta \gamma}}{3} \int_{-\infty}^{E} d\omega \int (dk) \,\times\\ 
             &\qquad\qquad\times   \tr\left(\tau_3 G_{\omega} \partial_{\alpha} G_{\omega}^{-1} G_{\omega} \partial_{\beta} G_{\omega}^{-1}G_{\omega} \partial_{\gamma} G_{\omega}^{-1}\right),\nonumber
     \end{align}
%\end{widetext}
where  $G=G_E$ is the SCBA Green function Eq.\eqref{eq:SCBAGreenFunction}, latin
indices take the values $i,j=1,2$ and the Greek indices the values $\alpha, \beta,
\gamma = \omega, 1, 2$. Referring for the derivation of these representations from Eq.\eqref{eq:TopologicalCouplingConstants} to Appendix \ref{smrckastreda}, we note that in the IQH context these integral representations  are known as the Smrcka-Streda \cite{smrcka1977transport} Hall coefficients, $\sigma_{xy}^{\mathrm{I}}=\theta_1/2\pi$, and $\sigma_{xy}^{\mathrm{II}}=\theta_2/2\pi$. These parameters describe the Fermi surface ($\sigma_{xy}^\mathrm{I}$) and thermodynamic ($\sigma_{xy}^{\mathrm{II}}$, note the integral over all energies below the Fermi surface) contribution to the Hall response $\sigma_{xy}=\sigma_{xy}^{\mathrm{I}}+\sigma_{xy}^{\mathrm{II}}$ of an electron gas subject to a magnetic field.

We may now numerically compute  the complex self energy $\Delta E+i\kappa$ in self-consistent Born approximation along the lines of  section \ref{ssub:disorder_average_and_stationary_phase_analysis}, and then do the integrals. This procedure  yields estimates for the topological angle, which however remain of limited accuracy. Deviations arise because of the reliance on the  SCBA, whose range of applicability is limited to $E \gg  \kappa$ and $\sigma_{xx} \gg 1 $. For the former condition, if we go outside this regime, the self energy $\kappa$ is affected by scattering processes technically described by diagrams with crossing impurity scattering and for the latter condition $\sigma_{xx} \sim \mathcal{O}(1)$ indicates close proximity to the quantum Hall critical point, where the approach discussed in this paper is no longer suitable. While we did not investigate the contributions from further scattering processes in quantitative detail, our comparison to exact diagonalization shows that we obtain reasonable agreement, including in regimes where the theory is past the region of parametric control. 
\begin{figure*}[ht]
	\includegraphics[width=0.32\linewidth]{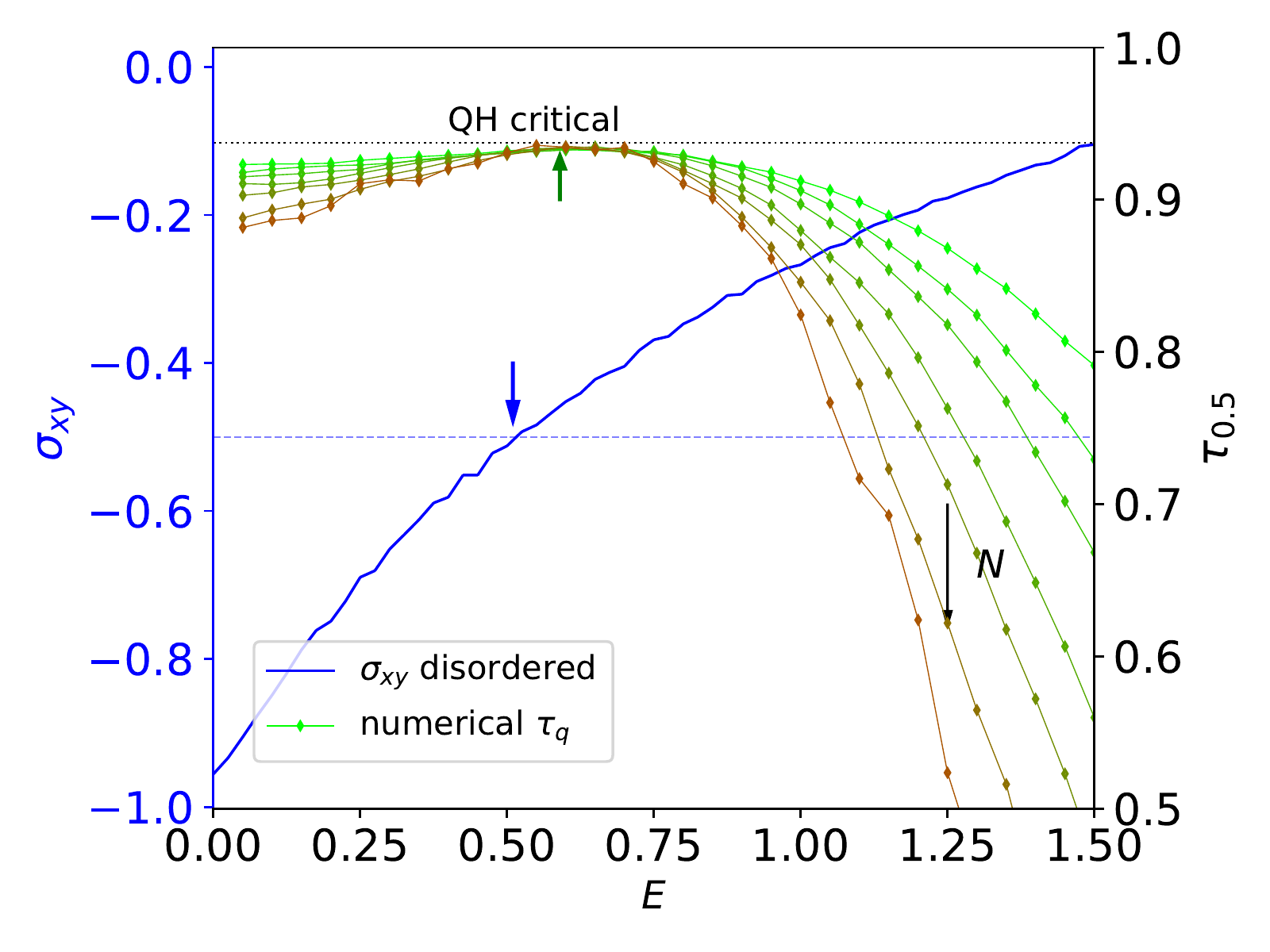}
	\includegraphics[width=0.32\linewidth]{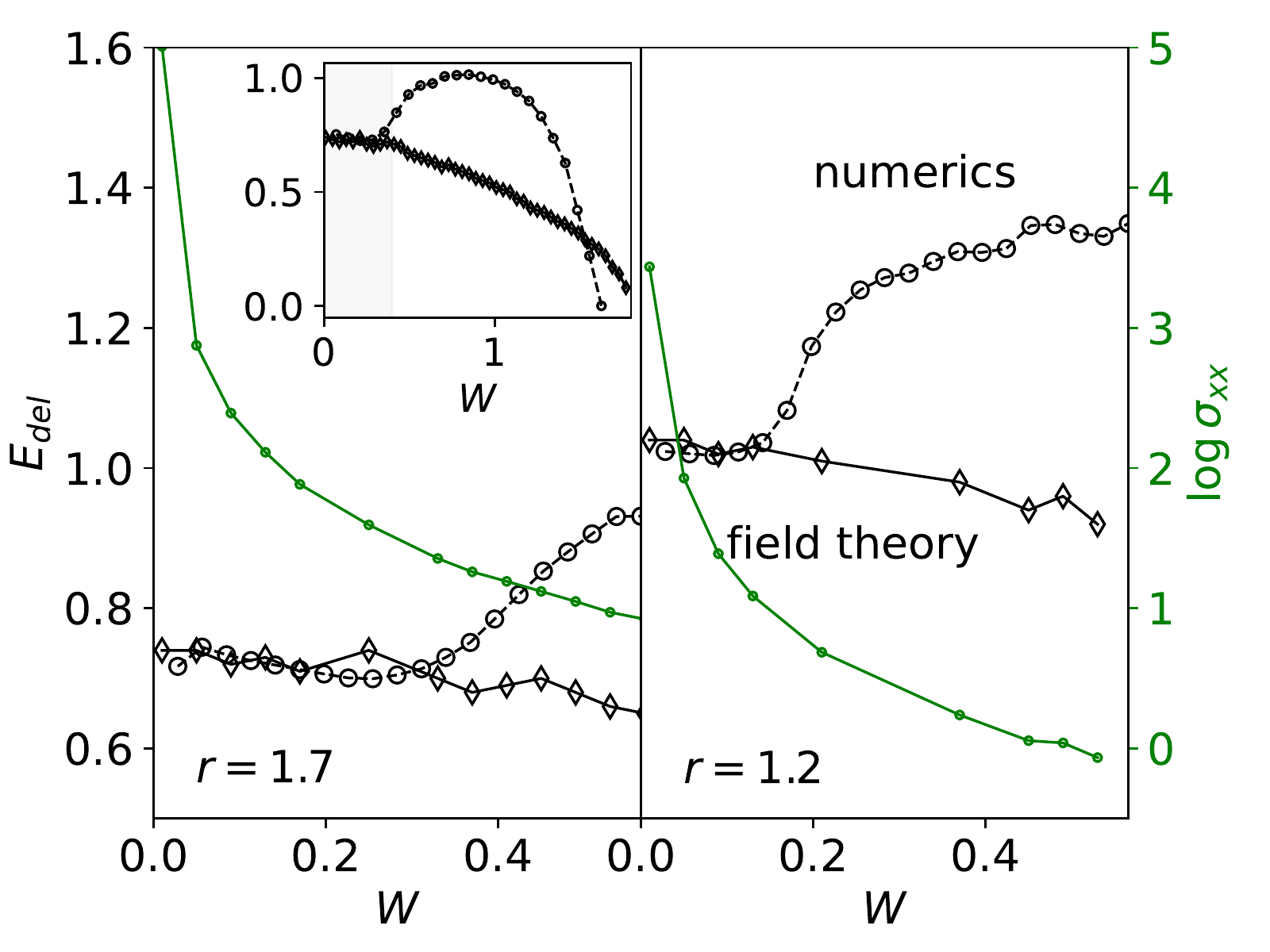}
	\includegraphics[width=0.32\linewidth]{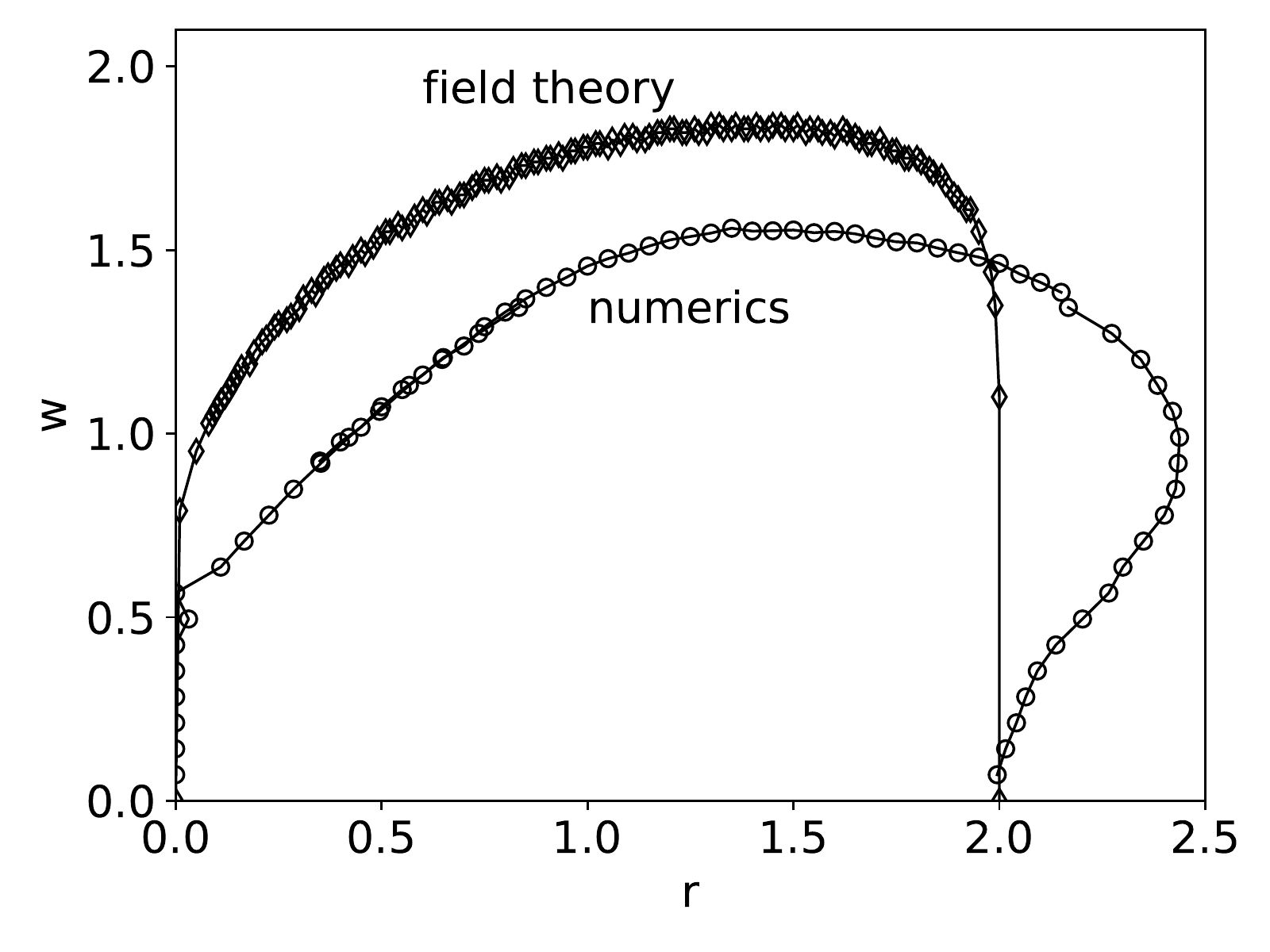}
	\caption{Analytical and numerical prediction for the position of the extended
	state. Left: the blue solid line shows the Hall conductivity calculated from
	Eq. \eqref{eq:TopAngleWeakDisorder}, for different energies, at
	$(r,W)=(1.2,1.45)$, while the blue dashed line shows the Hall conductivity for
	the clean Chern insulator at $r=1.2$. Criticality is associated with the value
	$\sigma_{xy}(E)=1/2\, \text{mod}\, 1$, cf. blue arrow. On the right axis the
	numerical results for the effective dimensions ${\tau}_q(E)$ \cite{note1} for
	different system sizes $N=64$ (green) up to $N=512$ (red) are shown (black
	arrow). The {dotted horizontal} line is the {dimension $\tau_{0.5}\approx
	0.94$} of the quantum Hall critical state. This condition is approximately met
	at the green arrow, which is reasonably, but not perfectly well aligned with
	the blue analytical marker. Center: prediction for the delocalized states
	taken from multifractal analysis and field theory at $r=1.7$ and $r=1.2$ as a
	function of disorder strength $W$. Additionally the analytically calculated
	$\sigma_{xx}$ is shown (green). The inset shows the behavior of the
	delocalized state calculated with both analytical and numerical approaches for
	the full topological phase at $r=1.7$. Right: phase diagram of the disordered
	Chern insulator in the $r$-$W$-plane at $E=0$, calculated analytically and
	numerically.
      }

	\label{fig:results}
\end{figure*}

\section{Comparison to exact diagonalization}
\label{sec:results}

In figure \ref{fig:results} we compare the results obtained from numerical
simulations (section \ref{sec:multifrac}) with the analytical predictions
(section \ref{sec:derivation_action}) for the position of the extended state in
the band of eigenstates. Numerically, we identify these states  by calculation
of the exponent  $\tau_{q}=2(1-q)+\Delta_q$, at quantum Hall criticality,
$\Delta_q =0.25\times  q(q-1)$, for the extremal value $q=0.5$, i.e.
$\Delta_\textrm{min}=\Delta_{q=0.5}\approx -0.06$, and $\tau_{0.5}\approx 0.94$.
 The leftmost panel shows the system size dependent
$\tau_{0.5}^{N}$ for  $r=1.2$, $W=1.45$ as a function of $E$. At the above value $\tau_{0.5}^N\approx \tau_{0.5}\approx 0.94$ (green arrow) the data becomes system size independent, signifying criticality with an exponent matching the quantum Hall expectation. 

To compare to the field theory predictions, we compute $\sigma_{xy}$ by
numerical evaluation of Eqs. (\ref{eq:sigmaxy}) for the same values of $r$ and
$W$ (blue curve). The crossing of the critical conductance $\sigma_{xy}=\frac{1}{2}$ is indicated by a blue arrow. The analytical and numerical predictions are not in perfect, but  in reasonable
agreement, given that there are no adjustable fitting
parameters. 

 The center panel shows the energy of the extended state at $r=1.7$ (left) and
	$r=1.2$ (right) as a function of the disorder strength $W$. The green curves
	show the analytically computed longitudinal conductance, where
	$\sigma_{xx}\gtrsim 1$ is  necessary for quantiative reliability of the field
	theory. As long as this condition is met, the analytical
	and numerical predictions for the value of the critical energy are in good
	agreement. At larger values of the disorder, $\sigma_{xy}$ is at its critical value, and $\sigma_{xx}=\mathcal{O}(1) $, so we are in proximity to the quantum critical point where the present theory is no longer applicable. In this regime, 
  the numerical prediction for the critical energy shows a transient increase (cf. inset) for which we do not have a good explanation. Eventually, the numerical and analytical value for the critical energy approach zero --- that they do so at roughly the same disorder concentration may be coincidental --- thus signalling the breakdown of the topological phase due to disorder. 
 
The right panel shows a cut through the critical surface at $E=0$. Inside the
lobe we have the Chern number $\mathrm{Ch}_c = -1$, outside it is vanishing.
Broadly speaking, we again observe semi-quantitative parameter free agreement between  field
theory and numerics. However, there are some qualitative features which the
former does not capture: Close to the clean critical value, $r=2$,
disorder stabilizes the topological phase in that the critical
value gets pushed upwards (the bulge visible in the numerical data.) This feature does not show in the field theoretical calculation. We suspect that this is due to the fact that we
are in parametric proximity  to a Dirac band closing at weak disorder and zero
energy. For such configurations, the SCBA approximation produces
incorrect estimates for self energies\cite{nersesyan1994}. However, a more detailed analysis of the latter beyond the SCBA approximation is beyond the scope of the present paper.

\section{Conclusions}
\label{sec:conclusions}

In the disordered Chern insulator, the spectral gap of the clean insulator is
replaced by a mobility gap: generic states inside the spectrum are Anderson
localized, thus preventing bulk hybridization  between the extended surface
states of the system. However, this feature cannot extend to all states: there must exist  bulk delocalized
states establishing contact between the surface bands somewhere up in the spectrum. However, these consistency arguments do not
tell us where the delocalized states lie in energy, nor what their critical
properties are. The study of these two questions was the subject of the present
paper. 

Describing the Chern insulator in terms of the three parameters  band energy,
$E$, effective disorder strength, $W$, and a parameter, $r$, controlling its
topological index $\nu(r)$, we applied a combination of analytical and numerical
methods to study the `critical surface' of delocalized states in the Chern insulator. For generic
parameter values, the critical states are buried deep in the band, meaning that
these analyses had to operate outside the regime where `Dirac' band
linearlizations are an option. Perhaps unexpectedly, this generalization turned
out to be a blessing, from various perspectives: The analytical derivation of an
effective field theory building on the full microscopic band structure was no
more difficult than the one starting from a lineraized spectrum. However, unlike
that one, it was not plagued by spurious ultraviolet divergences, and it produced
intuitive predictions for the identification of the critical states. Specifically, we found
that, at least for weak disorder, criticality was tied to the integral Eq.\eqref{eq:TopAngleWeakDisorder}: the
energy of critical states in the weakly disordered system is such that the
integrated Berry curvature of a all states above (or below) it equals $\pi$. The effective action describing the localization properties of these and of generic states was that of the quantum Hall insultator, confirming the expectation of bulk quantum Hall crtiticality in the system. However, for  generic points in the parameter space $(r,W,E)$ the weak disorder condition required for this description to be quantitatively reliable was violated, and quantitative errors ahad to be expected. 

To benchmark the quality of the analytical predictions, we analyzed the eigenstates of the system by numerical methods. Specifically, we computed the wave function scaling dimensions $\tau_{q}$,
Eqs.~\eqref{eq:InvParticipationRatio}, \eqref{eq_moments} to identify both  the
position of the extended states in parameter space, and their critical properties. This analysis confirmed the expectation of quantum Hall criticality, and for sufficiently weak disorder the results obtaind by the field theoretical computation. Outside that regime, the quality of the analytical predictions deteriorated, with errors up to $\mathcal{O}(1)$, but no parametric disagreement.    

The Chern insulator is one of the simplest topological insultators, and features as an effective building block for others. In view of the fact, that even  basic signatures of the (surface) criticality of disordered topological insulators remain mysterious\cite{PhysRevX.10.021025}, it is reassuring to have this basic system under control. We hope that the insights gained here may help in the solution of the more challenging problem of understanding the surface criticality of disordered three-dimensional topological insulators. 

\section{Acknowledgments}

J.D. thanks Ferdinand Evers and Martin Puschmann for useful discussions. M.M thanks Dmitry Bagrets for discussions. A.A.
and M.M. were funded by the Deutsche Forschungsgemeinschaft (DFG) Projektnummer
277101999 TRR~183 (project A03). J.D. acknowledges funding from the German
Academic Scholarship Foundation and computing time at the Gauss Centre for
Supercomputing via project pn72pa.

\appendix
\section{Self consistent Born approximation}
\label{sec:SCBA}
In this appendix we discuss more in depth the self consistent Born approximation and in general the role of the parameters $\kappa$ and $\Delta E$. The starting point is the 
self consistent Born equation for the matrix $A(x)$,

\begin{equation*}
	 A(x) = W^2 \tr \left(E+i\delta \tau_3-H-A \right)^{-1}(x,x).
\end{equation*} 
We propose a spatially homogeneous and matrix diagonal Ansatz of the form $\bar{A} = \Delta E + i \kappa \tau_3$. Plugging in the Ansatz into the previous expression, we obtain a self-consistent equation for both $\Delta E$ and $\kappa$.

\begin{equation*}
	\Delta E + i \kappa \tau_3 = \frac{W^2}{2} \int_{BZ} \frac{d^2k}{(2\pi)^2} \, \tr \left( \frac{1}{E-H(k)-\Delta E-i\kappa \tau_3}\right).
\end{equation*}

The real part of the previous equation, $\Delta E$, represents nothing more than an overall shift in the energy $E$ of the system. The imaginary part is the self energy due to impurity scattering which is to be identified with (2 times) the scattering rate off impurities and consequently defines another quantities of interest such as the elastic scattering time $\tau$ and the mean free path $\ell$.

\section{Derivation of the topological action} % (fold)
 \label{sec:derivation_of_the_topological_action}

We here derive Eq.\eqref{eq:TopologicalAction} for the topological action by explicit computation of the two contributing terms $S^{(1,2)}_\mathrm{top}$, i.e. the skew-derivative contributions to the first and second order gradient term in Eq.~\eqref{eq:actexpansion}.

\noindent $S_\mathrm{top}^{(2)}$:  The expanded representation of the second order term reads
\begin{widetext}
    \begin{align}
      S^{(2)}[Q]&=\frac{1}{2}\int dx(dk)\,\mathrm{tr}(D(i \kappa \tau_3 + v_\mu \sigma^\mu )F_{i}\Phi_{i} D(i \kappa \tau_3 + v_\nu \sigma^\nu )F_{\bar i}\Phi_{\bar i}) \cr
      &\to-\frac{1}{2}\int dx(dk)\,\mathrm{tr}(D(i \kappa \tau_3 + v_\mu \sigma^\mu )(\sigma_a \partial_i h_a \Phi_i) D(i \kappa \tau_3 + v_\nu \sigma^\nu ) (\sigma_b \partial_{\bar i} h_b \Phi_{\bar i})) \cr
      &= -i\kappa \int dx(dk)\,\mathrm{tr}(D \tau_3 (\sigma_a \partial_i h_a \Phi_i) D  (h_c \sigma_c)(\sigma_b \partial_{\bar i} h_b \Phi_{\bar i}) )  \cr 
      &= -2\kappa \epsilon_{abc} \int dx(dk)\,\mathrm{tr}(D \tau_3  \Phi_i D    \Phi_{\bar i} ) \partial_i h_a \partial_{\bar i} h_b h_c= -2\kappa \epsilon_{ij}\int dx(dk)\,\mathrm{tr}(D \tau_3  \Phi_i D    \Phi_{j})F_k,
    \end{align}
    \end{widetext}
    where $(dk)=dk_1 dk_2/(2\pi)^2$, the arrow indicates that we retain only derivative combinations $\partial_i\partial_{\bar i}$, $\bar i = (i+1) \mathrm{mod}\,2$, and we used the definition Eq.~\eqref{eq:Fk_def}.
 To process the integral over $k$, we decompose the matrices $D=D^+ P^++D^-
 P^-$, $P^s=\frac{1}{2}(1+s \tau_3)$ into advanced and retarded contributions  and
 note that only momentum integrals over denominators $D^+D^-$ of opposite causality
 are non-vanishing. In this way we arrive at
\begin{align}
\label{eq:I2Def}
    S^{(2)}_\mathrm{top}[Q]&= -\frac{\theta_1}{4\pi} \int dx \sum_s s \epsilon_{ij} \tr(P^s  \, \Phi_i \, P^{\bar s}  \Phi_j),  
\end{align} 
with the momentum integral $\theta_2$ defined in Eq.~\eqref{eq:TopologicalCouplingConstants}.
We finally use the first of the auxiliary relations 
\begin{align}
  \label{eq:PhivsII}
  -4 \epsilon_{ij}\sum_{s}\,\mathrm{tr}( s P^s \Phi_i P^{\bar s} \Phi_j)&= \mathcal{L}_\mathrm{top}(Q),\cr 
  4 \epsilon_{ij}\mathrm{tr}(\tau_3\partial_i \Phi_j)&=\mathcal{L}_\mathrm{top}(Q),
\end{align}
to obtain $S^{(2)}_\mathrm{top}$ as given in Eq.~\eqref{eq:TopologicalAction}.

\noindent $S_\mathrm{top}^{(2)}$: Being first order in derivatives, the contribution from the term $S^{(1)}$ naively
seems to vanish. To see that it does not, we play a trick first applied by Pruisken
in his analysis of the quantum Hall effect. Noting that the energy-dependent Green
function $G\equiv G_E$ can be written as $G_E=\int_E^\infty d\omega\, G_\omega^2$, we
represent the action as (in the same notation, $D_\omega\equiv D_{E=\omega}$)
\begin{widetext}
     \begin{align}
    S_\mathrm{top}^{(1)}[Q]&= -\int\limits_{E}^\infty d\omega \Tr(G_\omega F_i \Phi_iG_\omega)\to
    -\frac{i}{2} \int\limits_{E}^\infty d\omega \int dx (dk)\tr((\partial_jG_\omega) F_i (\partial_j\Phi_i)G_\omega- G_\omega F_i (\partial_j\Phi_i)\partial_jG_\omega)\cr 
    &=-\frac{i}{2} \int\limits_{E}^\infty d\omega \int dx (dk) \tr([(\partial_jG_\omega),G_\omega] F_i \partial_j\Phi_i)=-\frac{1}{2} \int_{E}^\infty d\omega\int dx (dk)\tr (D_{\omega}^2[\partial_j (h_a \sigma_a) ,h_b\sigma_b]  (\partial_i h_c \sigma_c) \partial_j\Phi_i)\cr
        &=-2i \int\limits_{E}^\infty d\omega \int dx (dk)\epsilon_{ij}\tr(D_{\omega}^2    \partial_j\Phi_i)F_k.
  \end{align} 
 \end{widetext} 
We now decompose the matrix $D$ again, and note that only the contribution
proportional to $\tau_3$ yields a non-vanishing trace, $D^2\rightarrow
\frac{1}{2}(D^{+2}-D^{-2})\tau_3$. As a result, we obtain 
\begin{align} \label{eq:I1Def}
  S_\mathrm{top}^{(1)}[Q]&= \frac{\theta_2}{4\pi}\int dx \,\epsilon_{ij}\,\mathrm{tr}(\tau_3 \partial_i\Phi_j),  
\end{align}
with $\theta_2$ given in \eqref{eq:TopologicalCouplingConstants}. 
In a final step, we use the second of the auxiliary relations \eqref{eq:PhivsII}
to arrive at the contribution $S^{(1)}$ to \eqref{eq:TopologicalAction}.

\section{Derivation of the gradient action} % (fold)
\label{sec:derivation_of_the_gradient_action}

We here derive Eq.~\eqref{eq:GradientAction} by inspection of the two terms $S^{(1,2)}$ in the formal gradient expansion Eq.~\eqref{eq:actexpansion}.

\noindent  \emph{$S^{(1)}_\mathrm{grad}$:} Filtering symmetric derivative combinations from the explicit representation of the second order expansion we obtain
\begin{widetext}
\begin{align*}
        S^{(2)}_{\mathrm{grad}}[Q] &=
        \frac{1}{2}\int dx (dk)\tr(D(i\kappa \tau_3+v_\mu\sigma^\mu)F_i \Phi_i D(i\kappa \tau_3+v_\nu\sigma^\nu)F_{j}\Phi_{j})\\
        &\to  -\frac{1}{2}   \int dx \, (dk) \tr\left(-\kappa ^2 D \tau_3 \sigma_{\nu} \Phi_{i} D \tau_3  \sigma_{\lambda} \Phi_i   + D v_{\mu} \sigma^{\mu} \sigma_{\nu} \Phi_{i} D v_{\rho} \sigma^{\rho} \sigma_{\lambda} \Phi_{i}    +2i E \kappa\,  D \tau_3  \sigma_{\nu} \Phi_i D \sigma_{\lambda} \Phi_i \right)\partial_i v_{\nu} \partial_i v_{\lambda}  \\ 
        &=  -\int dx (dk)\sum_a \tr \left(-\kappa^2\,   \tau_3 D \Phi_i \tau_3 D \Phi_i + (E^2- \epsilon^2+ 2 h_a^2 )  D \Phi_i D \Phi_i   +  2i E \kappa\, D \tau_3 \Phi_i D \Phi_i \right) \partial_i h_{a} \partial_i h_{a},
\end{align*}
\end{widetext}
where  "$\to$" indicates that we retain only derivatives with identical $i$-index,
and in the second equality traced over Pauli matrices. To compute the $k$-integrals,
we again decompose $D = D^+P^+ + D^-P^-$. The product of two $D$'s then leads to
terms $D^s D^{s'}$ of equal and opposite causal index $s,s'$, which need to be
considered separately.

Using the auxiliary relations 
\begin{align*}
   & \sum_{ s}  \tr \left( \Phi_i \tau_3^n P^s \Phi_i \tau_3^m P^{-s} \right) = \frac{1}{4}\tr \left( \partial_i Q \partial_i Q \right)\times \cr 
    &\qquad\times \left\{\begin{array}{ll}
        -1&(n,m)=(0,0),\cr 
        1&(n,m)=(1,1),\cr 
        0&(n,m)=(0,1),(1,0)
    \end{array} \right. ,
\end{align*}
and
\begin{align*}
    &\sum_{ s}  \tr \left( \Phi_i \tau_3^n P^{s}\Phi_i \tau_3^m P^s \right)f_s =\cr 
   &\quad = -\sum_s s^{n+m}\tr\left( \frac{1}{4}  \partial_i Q \partial_i Q  -  P^s\Phi_i^2\right)f_s,
\end{align*}
where $f_s$ is arbitrary, it is straightforward to obtain
\begin{align}
    S^{(2)}_{\mathrm{grad}}[Q] = (I_{+} + I_{-}+I_{+-}) \int dx \tr \left( \partial_i Q \partial_i Q \right) + S_{\mathrm{A}},
\end{align}
with $S_\mathrm{A}= 4 \sum_{s} I_{s} \int dx \tr(P^s \Phi_i^2)$, and the coefficients defined in Eq.~\eqref{eq:GradientCouplingConstants}

\noindent \emph{$S^{(1)}_\mathrm{grad}$:} The terms with equal indices, $J_{ii}$ in Eq.~\eqref{eq:actexpansion} yield a term 
\begin{align*}
    S^{(1)}_{\mathrm{grad}} &= \frac{1}{2} \Tr (G J_{ii}\Phi_i^2), \\
    &= \frac{1}{2} \int dx (dk) \sum_{s} \tr \left( G^s P^s \partial_i^2 v_{\mu}\sigma_{\mu} \Phi_i^2 \right), \\
    &= \frac{1}{2} \int dx (dk) \sum_{s} \tr \left(G^s \partial_i v_{\nu} \sigma_{\nu} G^s \partial_i v_{\mu} \sigma_{\mu} P^s \Phi_i^2\right), \\
    &= -S_\mathrm{A}
\end{align*}
where we integrated by parts and used that $\partial_i G^s = -G^s \partial_i
h_{\nu}\sigma_{\nu}G^s$. We conclude that the anomalous terms, $S_\mathrm{A}$ cancel out and arrive at the full gradient action Eq.~\eqref{eq:GradientAction}.

\section{Derivation of Eq. (\ref{eq:intI})} % (fold)
  \label{sec:derivation_of_eq_}
 
  In this appendix we take a closer look at the derivation of equation (\ref{eq:intI}). The first thing to notice is that $I$ can be written as the following,
  \begin{widetext}
  	\begin{align*}
  		I =& \frac{1}{4} \sum_{a} \int (dk) ((E+i\kappa)D^{+}+(E-i\kappa)D^{-})^2 + (2h_a^2-\epsilon^2)(D^{+}+D^{-})^2 (\partial_i h_a \partial_i h_a), \\
  		=& \sum_{a} \int (dk) \frac{E^2(E^2+\kappa^2-\epsilon^2)^2+(2h_a^2-\epsilon^2)(E^2-\kappa^2-\epsilon^2)^2}{((E^2-\kappa^2-\epsilon^2)^2+4E^2\kappa^2)^2} (\partial_i h_a \partial_i h_a), \\
  		=& \sum_{a} \int (dk) \frac{E^2(E^2+\kappa^2-\epsilon^2)^2+(2h_a^2-\epsilon^2)(E^2-\kappa^2-\epsilon^2)^2}{((E^2-\kappa^2-\epsilon^2)^2+4E^2\kappa^2)(2|E|\kappa)}\left( \frac{2|E|\kappa}{(E^2-\kappa^2-\epsilon^2)^2+4E^2\kappa^2} \right) (\partial_i h_a \partial_i h_a).
  	\end{align*}
  \end{widetext}
  To make further progress we take the limit when $E \kappa \to 0$ resulting in,
  
  \begin{align*}
  	I =& \, \pi \sum_{a} \int (dk) \, \frac{E^2-\epsilon^2+2h_a^2}{2|E|\kappa} (\partial_i h_a \partial_i h_a) \delta(E^2-\epsilon^2), \\
  	=& \, \pi \int (dk) \frac{\sum_{a} h_a^2 (\partial_i h_a \partial_i h_a)}{|E| \kappa} \delta(E^2-\epsilon^2).
  \end{align*}
  
  At this point we focus in the low energy regime, where we can take the Dirac approximation, resulting in,
  
  \begin{equation*}
  	I = \frac{E^2-m^2}{2|E|\kappa} \Theta(E^2-m^2),
  \end{equation*}

with $m=r-c$ and $c=2,0,-2$ depending on the Dirac cone around which we approximate.

\section{Derivation of the Smrcka-Streda coefficients}
\label{smrckastreda} 

In this appendix we show the relation between the equations \ref{eq:TopologicalCouplingConstants} and the equations (\ref{eq:sigmaxy}). More precisely, we want to show that $\theta_1 = 2\pi \sigma_{xy}^{I} $ and $\theta_2 = 2\pi \sigma_{xy}^{II}$.

\begin{widetext}
	\begin{align}
		\begin{split}
			\sigma_{xy}^{I} &= -\frac{i}{16\pi^2} \int dk \, \epsilon_{ij} \tr\left(\tau_2 G_E \partial_i G_E^{-1}\tau_1 G_E \partial_j G_E^{-1}\right), \\
			&= \frac{1}{8 \pi^2} \sum_{s} \int dk \, s \tr((G_E\partial_1G_E^{-1})_s(G_E\partial_1G_E^{-1})_{-s}), \\
			&= \frac{1}{8 \pi^2} \sum_{s} \int dk \, s \tr \left((E+is\kappa+h_a\sigma_a)(\partial_1 h_b \sigma_b)(E-is\kappa+h_c\sigma_c)(\partial_2 h_d \sigma_d)\right)D^{s}D^{-s},\\
			&= -\frac{i \kappa}{2\pi^2} \int dk \tr(\sigma_a \sigma_b \sigma_c)h_a \partial_1 h_b \partial_2 h_c D^{+}D_{-} , \\
			&= \frac{\kappa}{\pi^2} \int dk \, D^{+}D^{-}F_{k}, \\
			&= \theta_1/(2\pi), \\
		\sigma_{xy}^{II} &= \frac{1}{24\pi^2} \int_{-\infty}^{E} \int d\omega dk \, \epsilon_{\alpha \beta \gamma} \tr\left(\tau_3 G_{\omega} \partial_{\alpha} G_{\omega}^{-1} G_{\omega} \partial_{\beta} G_{\omega}^{-1}G_{\omega} \partial_{\gamma} G_{\omega}^{-1}\right), \\
		&= \frac{1}{8 \pi^2} \sum_s \int_{-\infty}^{E} d\omega \int  dk \, s \tr \left( ((G_\omega \partial_1 G_\omega^{-1})_s (G_\omega \partial_2 G_\omega^{-1}) - (G_\omega \partial_1 G_\omega^{-1})_s (G_\omega \partial_2 G_\omega^{-1})G_\omega)\right), \\
		&= \frac{1}{8 \pi^2} \sum_s \int_{-\infty}^{E} d\omega \int  dk \, s \tr \left( [(\partial_1 G_\omega)_s,(G_\omega)_s]\partial_2 (G_\omega^{-1})_s \right),\\
		&= \frac{1}{8 \pi^2} \sum_s \int_{-\infty}^{E} d\omega \int  dk \, s \tr \left( [(\omega + is\kappa+h_a\sigma_a),\partial_1 h_b \sigma_b] \partial_2 h_c \sigma_c   \right) D_s^{2}, \\
		&= \frac{1}{4 \pi^2} \int_{-\infty}^{E} d\omega \int  dk \, \tr(\sigma_a \sigma_b \sigma_c)h_a \partial_1 h_b \partial_2 h_c (D_{+}^2-D_{-}^2), \\
		&= \frac{i}{2\pi^2} \int_{-\infty}^{E} d\omega \int  dk \, F_k ((D_{+}^2-D_{-}^2)),\\
		&=\theta_2/(2\pi).		
	\end{split}
   \end{align}
\end{widetext}
 
 \section{Details of the multifractal analysis}
 \label{app_num}
 This section provides supportive data to show that the critical lines shown in Fig. \ref{fig:results} can be identified with quantum Hall criticality. 
 \subsection{Distribution functions of the moments}
\begin{figure}[h!]
 	\includegraphics[width=\linewidth]{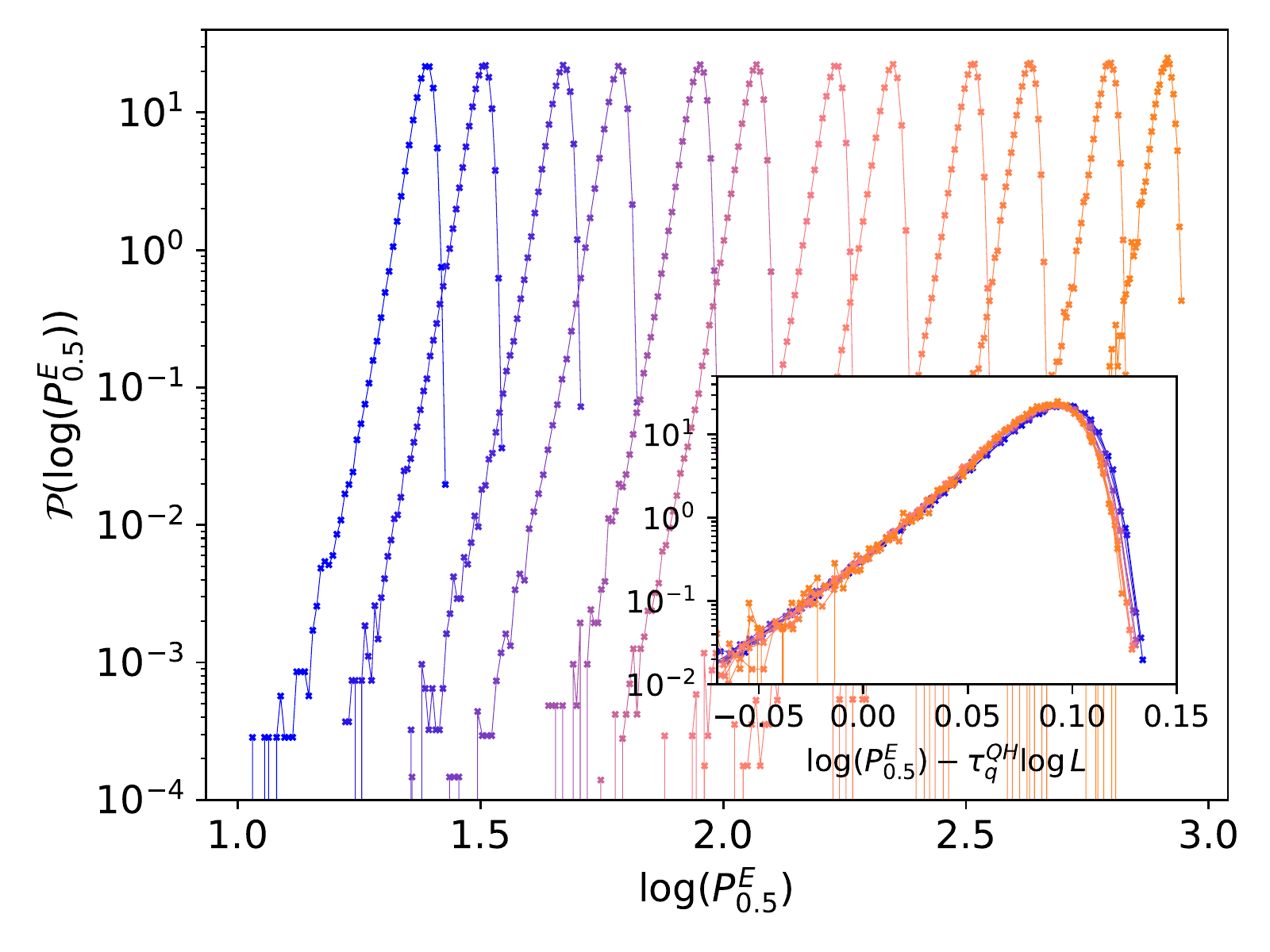}
 	\caption{Distribution functions of $P_{0.5}$ at $E=0.59, W=1.45, r=1.2$. According to the field theoretical calculations a quantum Hall critical states should exist here. The distribution functions correspond to different system sizes, $L=16,24,32,48,64,96,128,192,256,384,512,768,1024$ (blue to red). Not only their mean, which is relevant for the scaling, but the full distribution function collapses to a power law governed by the quantum Hall critical dimension $\tau_{0.5}^{QH}\approx 0.94$ (inset). }
 	\label{f_app1}
 \end{figure}

 Fig. \ref{f_app1} shows the distribution functions for the critical point associated with the delocalised state found analytically in Fig. \ref{fig:results} (left). The shape of the distribution function as well as their mean converge to a power law scaling governed by the effective dimension of the quantum Hall effect, known from other numerical works. \cite{evers2008multifractality} This can be most convincingly seen in the collapsed curves in the inset of Fig. \ref{f_app1}. 
 
 \subsection{Extrapolation of the critical exponents}

 In order to find the critical point at which we can compare to quantum Hall criticality in the first place we calculate $\tilde \tau_q(E,W,r)$ at $q=0.5$ as described in the main text for all available system sizes, and find their local maxima.

 A cut through phase space close to the critical point $(E^*,W^*,r^*)\approx(0.6,1.45,1.2)$ is shown in Fig. \ref{f_app2}.

  The extrapolation of the critical properties is done by fitting the curves using the model 
 \begin{equation}
 	\tilde{\tau}_q=\tilde\tau^c_q+\frac{1}{2}\,\tilde{\tau}_{q}^{c\,\prime\prime}\,(W-W^c)^2+\beta^c(W-W^c)^3,
 	\label{eq_app_fit_para}
 \end{equation}
 where the superscript $c$ denotes the fitted values at criticality for a given system size $N$, and $\alpha^c$ the opening angle of the parabolic part of the curve, from which the curvature, and the localization length exponent can be extracted. We use a third order model to account for asymmetries in the curves. 
 
 \begin{figure}
 	\centering 
 	\includegraphics[width=\linewidth]{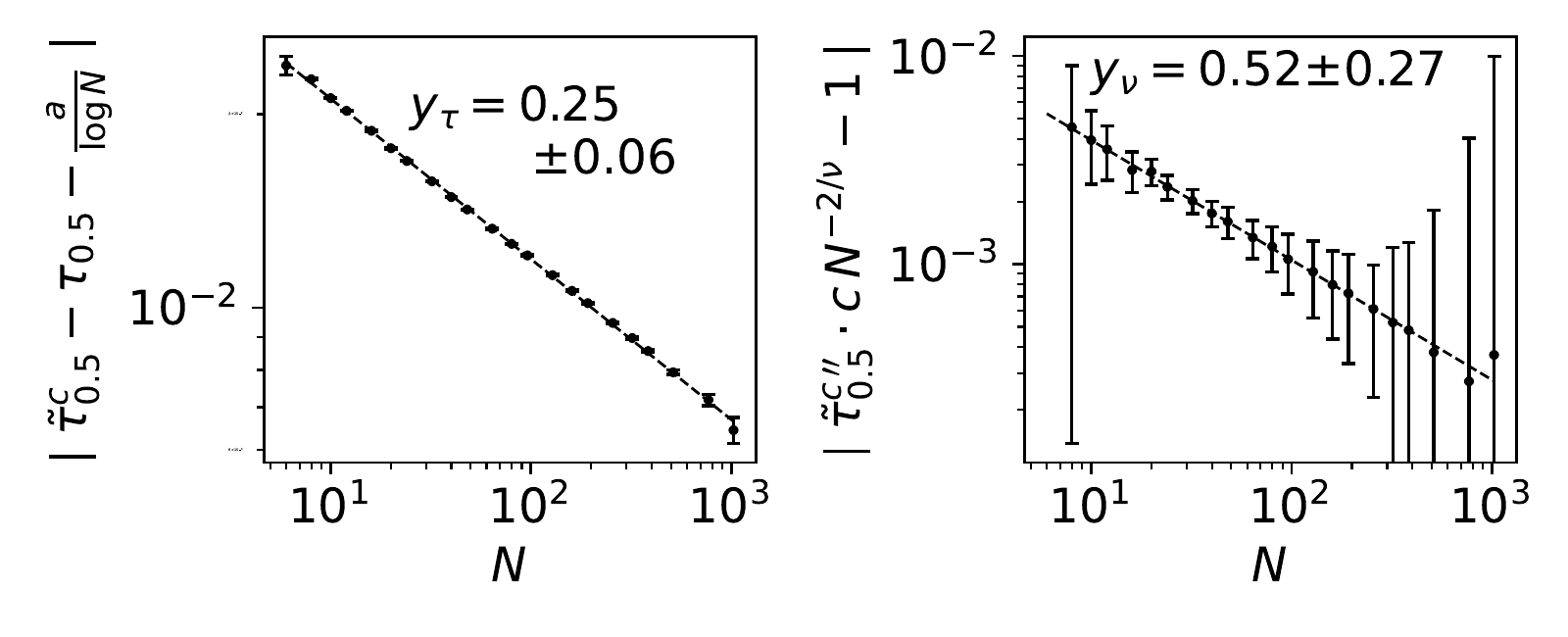}
 	\caption{(left) The irrelevant system size scaling of the fitted maxima without the
 		logarithmic part originating from the simplified definition of $\tilde{\tau}_q$.
 		(right) irrelevant exponent $y_\nu$ from
 		the fitting function in Eq. \eqref{fit_nu}. The irrelevant exponents quantify the finite size corrections associated with the exponents extracted in Fig. \ref{f_app2}. }
 \end{figure}
 
 In Fig. \ref{f_app3} we present a more detailed analysis of the system size scaling of the data for the effective dimension in Fig. \ref{f_app2}. 
 Since we approximate the true effective dimension $\tau_q$ by the maxima of $\tilde{\tau}_q$, we need to extrapolate to $N\to\infty$. This is done by the fitting function
 \begin{equation}
 	\tilde{\tau}^c_q(N)=\tau_q+\frac{a}{\log N}+b\,N^{-y_{\tau}},
 	\label{fit_tau}
 \end{equation}
where the fitting parameter $a$ corresponds to the prefactor of the system size scaling of the wave function moments in Eq. \eqref{eq_moments} and $y_{\tau}$ approximates the exponent of an irrelevant scaling correction.
 
Additionally, we show the localization length exponent $\nu$ extracted from the curvature of the curves in Fig. \ref{f_app2}. The latter is supposed to scale as 
\begin{equation}
	\tilde{\tau}^{c\,\prime\prime}_q=c\,N^{2/\nu}\, (1+d\,N^{-y_{\nu}}),
	\label{fit_nu}
\end{equation}
 where again $y_\nu$ denotes the corresponding irrelevant exponent. 
 We find $\nu=2.73\pm0.16$ and $y_{\nu}=0.52\pm0.27$ as described in the main text. 
 
\bibliography{paperbibl.bib}

\begin{thebibliography}{32}
\bibitem [{note1()}]{note1}%
  \BibitemOpen
  \bibinfo {note} {Here for a better visualization of the convergence, of the
  value of the effective dimension itself we calculate
  $\tau_q^N=(\log(P_q^N)-\log(P_q^{N/2}))/(\log N -\log N/2)$. This avoids the
  logarithmic correction in Eq. \eqref{fit_tau}}\BibitemShut {NoStop}%
\end{thebibliography}

\end{document}